\documentclass[epj]{svjour}
\usepackage{amsfonts,amssymb}
\usepackage{natbib}
\usepackage{graphicx}

\newcommand{\mean}[1]{\left\langle #1 \right\rangle}

\begin{document}
\title{Universality in movie rating distributions}
\author{Jan Lorenz 
}                     
%
%
\institute{Chair of Systems Design, ETH  Z\"urich, Kreuzplatz 5, 8032 Z\"urich,
Switzerland, jalorenz@ethz.ch}
\date{Received: date / Revised version: date}
%
\abstract{
In this paper histograms of user ratings for movies
($1\bigstar,\dots,10\bigstar$) are analysed. The evolving stabilised shapes of
histograms follow the rule that all are either double- or triple-peaked.
Moreover, at most one peak can
be on the central bins $2\bigstar,\dots,9\bigstar$ and the distribution in these
bins looks smooth `Gaussian-like' while changes at the extremes ($1\bigstar$ and 
$10\bigstar$) often look
abrupt. It is shown that this is well approximated under the assumption that
histograms are
confined and discretised probability density functions of L\'evy skew
$\alpha$-stable distributions. These distributions are the only stable
distributions which could emerge due to a generalized central limit theorem from
averaging of various independent random variables as which one can see the
initial
opinions of users. Averaging is also an appropriate assumption about the social
process which underlies the process of continuous opinion formation.
Surprisingly, not the normal distribution achieves the best fit over histograms
obseved on the web, but distributions with fat tails which decay as power-laws
with exponent $-(1+\alpha)$ ($\alpha=\frac{4}{3}$). The scale and skewness
parameters of the L\'evy
skew $\alpha$-stable distributions seem to depend on the deviation from an
average movie (with mean about $7.6\bigstar$). The histogram of such an average
movie
has no skewness and is the most narrow one. If a movie deviates from average the
distribution gets broader and skew. The skewness pronounces the deviation.
This is used to construct a one parameter fit which gives some evidence of
universality in processes of continuous opinion dynamics about taste. %
\PACS{
      {89.20.Hh}{World Wide Web, Internet}   \and
      {89.75.Da}{Systems obeying scaling laws}
     } 
} 
\maketitle
\section{Introduction}

Are there universal laws underlying the dynamics of opinion formation?

Understanding opinion formation is tackled classically by social
psychologists and sociologists with experiments
(see e.g. \cite{Asch1955Opinionsandsocial, Lorge.Fox.ea1958surveyofstudies,
Sherif.Hovland1961SocialJudgmentAssimilation,
Forgas1977Polarisationandmoderation,
Friedkin1999ChoiceShiftand,Salganik.Dodds.ea2006ExperimentalStudyof,
Mason.Conrey.ea2007SituatingSocialInfluence}),
but also by the social simulation (see e.g.
\cite{Deffuant.Neau.ea2000MixingBeliefsamong,
Hegselmann.Krause2002OpinionDynamicsand, Urbig2003AttitudeDynamicswith,
  Deffuant.Neau.ea2002HowCanExtremism,
Jager.Amblard2004UniformityBipolarizationand,
Salzarulo2006ContinuousOpinionDynamics,
 Baldassarri.Bearman2007DynamicsofPolitical})
and sociophysics (see surveys
\cite{Stauffer2005SociophysicssimulationsII,
Lorenz2007ContinuousOpinionDynamics,
Castellano.Fortunato.ea2007Statisticalphysicsof})
communities. Often studies are either empirical but on small experimental
samples or contrary they analyse models analytically or by simulation but
without empirical validation. Both restricts the possibility to draw
conclusions on universality in real world opinion formation. This is to a
large extent due to the difficulties in getting large scale data on human
opinions. But this situation changes rapidly nowadays thanks to the world
wide web. The existence of rating modules is almost
ubiquitous. (In the meantime the ubiquity of ratings has raised
  the question how to standardise rating modules
  \cite{Turnbull2007Ratingvoting}.)

This paper is an attempt to exploit rating data to extract universal
properties in opinion formation processes. Specifically, the focus here
is on opinions about the quality of movies, as expressed by users on movie
rating sites. Ratings stand as a proxy
for any opinion related to taste which is one-dimensional and of a
continuous nature (`continuous' means expressible as a real
  number and also gradually adjustable (at least to some
  extent)). Apparently, possible user ratings are discrete
($1\bigstar$(awful), $\dots$, $10\bigstar$(excellent)), but the continuous
nature (in the sense of ordered numbers) is also obvious.

Thus, this paper is not about discrete opinion dynamics without a
continuous nature (like e.g. with respect to decision: `yes' or `no') as
often studied in physics because of the analogy to spin systems. This
paper is also not on multidimensional many-faceted opinions (as
e.g. \cite{Lorenz2008ManagingComplexityInsights,
  Baldassarri.Bearman2007DynamicsofPolitical}) but on issues which are
broken down to one variable: the quality of a movie. It is also
important to distinguish the type of opinion. Movie ratings are about taste. 
There is no true value as for example in issues of
fact-finding about an unknown quantity. Further on, there is no real
physical constraint for opinions. It is always possible to like a movie
more than someone else. This is for example not the case in opinions
about budgeting in the political realm, where opinions have to be within
certain bounds. Finally, taste differs from issues about negotiations
where there is a clear incentive of agreeing on a common value (as e.g. for
prices in trade, or forming a politcal party in political issues). In issues of
taste there is
nevertheless a weaker force to adjust towards the opinions of peers, e.g. for
normative reasons (`I'd like to like what my peers like.'). But there might also
be a force to adjust away from the opinions
of others to pronounce individuality.

User ratings on the world wide web have already been subject of
research. Dellarocas \cite{Dellarocas2003DigitizationofWord} sketches their
role for digitising Word-of-Mouth (with the main focus on reputation
mechanisms). Ratings play a key role in some recommendation algorithms, see Goldberg \emph{et al}
\cite{Goldberg.Roeder.ea2001EigentasteConstantTime}, Cheung \emph{et al} \cite{Cheung.Kwok.ea2003Miningcustomerproduct}, and Umayarov \emph{et al}
 \cite{Umyarov.Tuzhilin2007Leveragingaggregateratings} which work by comparing
the rating profiles of different users. They also play a role in a recent
method of pricing an option on movie revenues, see Chance \emph{et al}
\cite{Chance.Hillebrand.ea2008PricingOptionRevenue}.
Salganik \emph{et al} \cite{Salganik.Dodds.ea2006ExperimentalStudyof}
study the emerging popularity of songs measured by downloads under the
impact of the visibility of the number of downloads. They used ratings
to check if liking corresponds to downloads, which is the case. But which
movie gets popular is to some extend arbitrary.

Jiang and Chen \cite{Jiang.Chen2007EconomicAnalysisof} argue economically that
the
implementation of online rating systems can enhance consumer surplus,
vendor profitability and social welfare. But they also argue, that this
could work better in a monopolistic market than a duopolistic market.

Cosley \emph{et al} \citet{Cosley.Lam.ea2003Isseeingbelieving} checked how users
re-rate
movies especially if they are confronted with a prediction of the quality
(like the mean of other ratings). They found a tendency to adjust towards
the presented prediction. They also show that users rate quite
consistently when they re-rate on other scales (like $5\bigstar$ compared to
$10\bigstar$).

Li and Hitt \cite{Li.Hitt2004SelfSelectionand} analysed the time
evolution of the user reviews arriving. (A review is a text but it is accompanied by
a rating is assigned by the writer.) They present an economic model where the
utility of a product
for a user is determined by individual search attributes which are
known before purchase and individual quality which can only be checked
after purchase. Both attributes are heterogeneous across the population
and purchasing decisions are made with respect to expected
quality. Expectations can be influenced by user reviews. Positive reviews
of early adopters produce high average ratings and thus too high expected
quality. This triggers purchases of other
consumers which then get disappointed and write bad reviews. If
individual search attributes towards a product are positively correlated
with individual quality then this may imply a declining trend of
reviews. This is called positive self selection bias. Negative
correlations imply negative self selection bias and thus an increasing
trend of reviews. These trends are confirmed empirically by book review
data on \texttt{amazon.com} with the majority of products (70\%) showing
positive self selection.

The phenomenon of declining average votes has been explained in a
different way in \cite{Wu.Huberman2008PublicDiscoursein}. They argue from
the point of view of the writer in front of a computer. Writing a
review is costly (in terms of time) and writers want to impact the
average vote. While the average vote over all books is more positive one
can only make a difference with a negative review, so writers with a
positive attitude hesitate to write a review. (If there are already a lot,
so why write another?) They also emphasize
that
internet reviews do not show a group polarization effect which is known
to appear in small groups discussing in the same physical
room\cite{Friedkin1999ChoiceShiftand}. 

There are few studies on characterising the empirical distributions of
ratings. In \cite{Dellarocas.Narayan2006StatisticalMeasureof} histograms of
user-ratings 
(on $5\bigstar$-scale in \texttt{movies.yahoo.com}) are characterised as
U-shaped, while professional critics have a single-peaked usage of the
votes (peak is at 4). Other studies concentrate either on user profile
comparison or only on the average vote and how it could impact further
votes and sales. In models idiosyncratic opinions are very often thought
to be normal distributed
\cite{Borghesi.Bouchaud2007Ofsongsand,Gu.Lin2006DynamicsOFOnline,
Umyarov.Tuzhilin2007Leveragingaggregateratings}. In
the model of \cite{Li.Hitt2004SelfSelectionand} the beta distribution is
used which lives on a bounded interval.

Normal distribution, Beta distribution, and U-shape all do not coincide
with the observation of rating histograms studied which are
very often triple peaked. In the following, the idea is introduced that
a rating of a user is derived from an originally continuous opinion from
the whole real axis. The opinio becomes a rating by discretising and confinig it
to the ratings scale. 
Further on, we assume that user's original opinions when it comes to rating are
already arithmetic averages of the expressed
opinions of peers, opinions of professional critics and possibly the
existing average (similar to the approach in
\cite{Gu.Lin2006DynamicsOFOnline}). 
This implies that limit theorems for sums of random variables play a role. 

\section{Empirical rating distributions and a simple model}

The aim of this paper is to characterise the distribution of ratings
towards a certain movie when the rating histogram contains a lot of ratings.
For a first analysis of the question some histograms of movie rating have been
collected \cite{dataset}

\begin{figure*}[htbp]
  \centering
\includegraphics[width=0.45\textwidth]{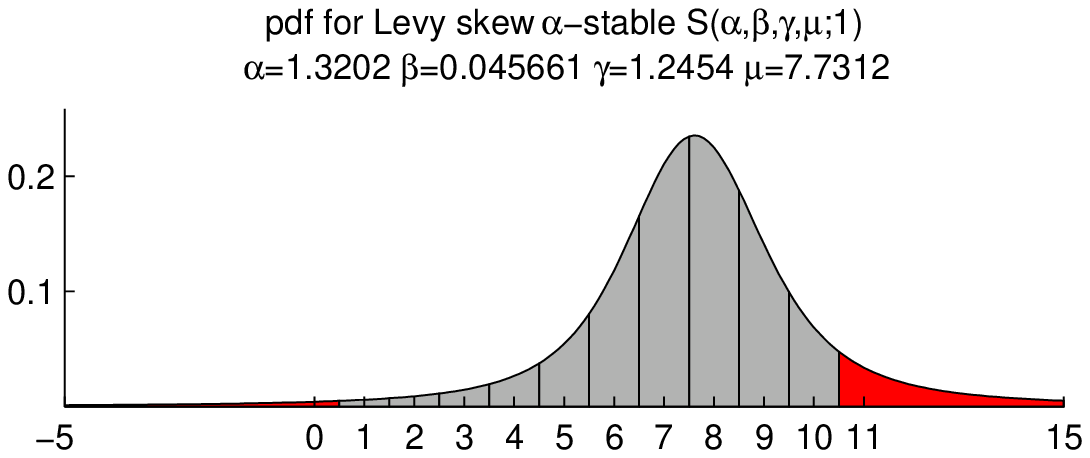}
\includegraphics[width=0.45\textwidth]{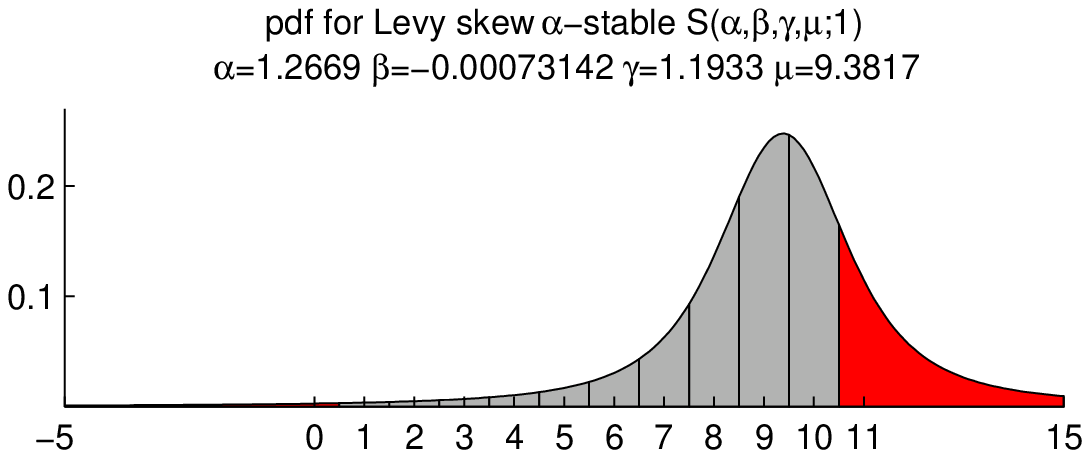}
\includegraphics[width=0.45\textwidth]{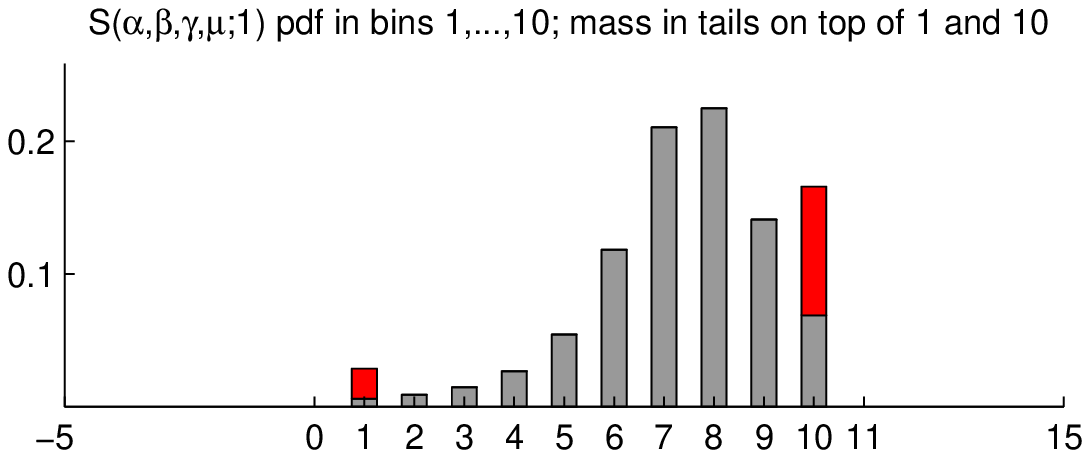}
\includegraphics[width=0.45\textwidth]{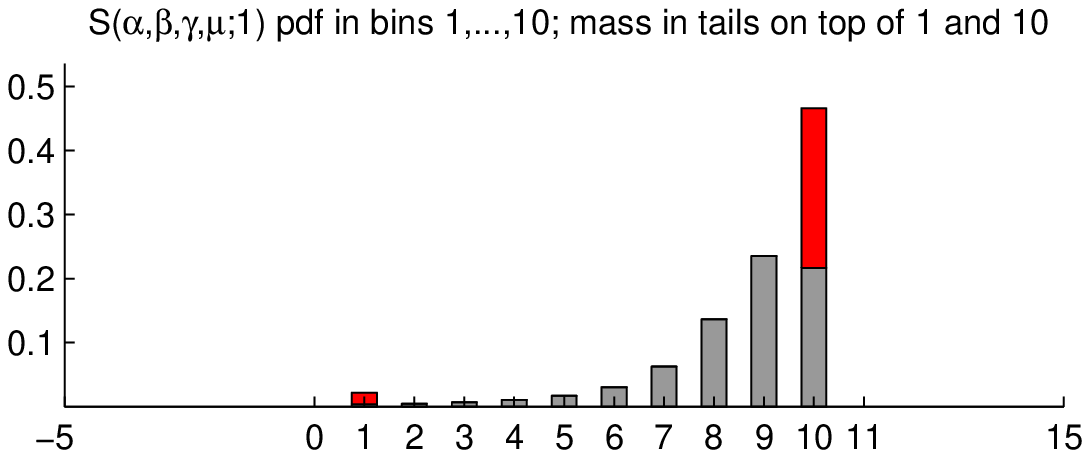}
\includegraphics[width=0.45\textwidth]{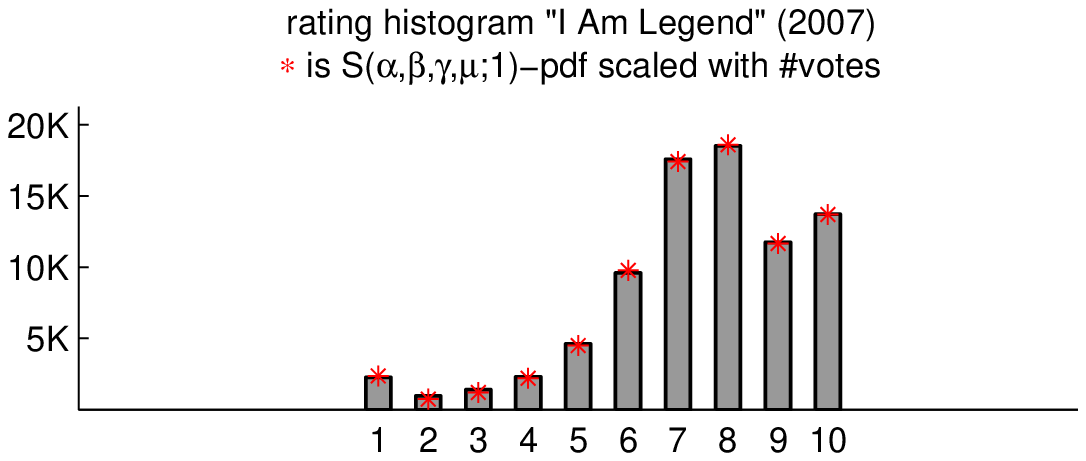}
\includegraphics[width=0.45\textwidth]{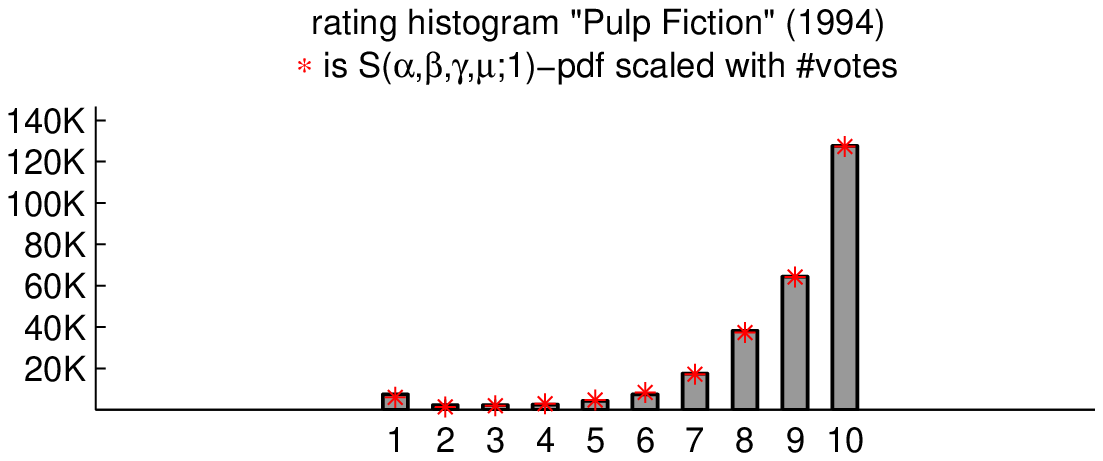}
  \caption{Explanation of confined L\'evy skew $\alpha$-stable distributions transformed to rating histograms. Shown
    are best fits for two examples movies. }
  \label{fig:figExplainRatingLevyskewalpha}
\end{figure*}

A brief inspection of a couple of histograms reveals
the following picture: Almost every histogram has either two or three
peaks. (A `peak' is a bin where all neighbour bins are less in
  size. It is a local maximum (or mode) of the probability mass function
  of the distribution.) In the case of two peaks at least one is at
1$\bigstar$ or 10$\bigstar.$. In the case of three peaks one is at
1$\bigstar$ and one at 10$\bigstar$. The histogram at the central bins
$2\bigstar,\dots,9\bigstar$ has a `Gaussian-function like' shape with a
peak and exponentially looking decay. This gives rise to the idea that
the histogram is a discretised version of a probability density function
on the real axis which is confined to the interval of possible ratings. Specifically,
we consider the opinion about a movie from cinemagoers to be a
real-valued random variable which is somehow distributed. When it comes
to assign stars the voter has to discretise her opinion
to the bins $1\bigstar,\dots,10\bigstar$. Naturally, the voter would
discretise according to the intervals $]-\infty,1.5], ]1.5,2.5], \dots ,
]8.5,9.5],]9.5,+\infty[$. If all voters draw their vote from the same
distribution the histogram will have bins with masses proportional to
the integrals of the probability density function (pdf) of that distribution over the
above intervals. Figure \ref{fig:figExplainRatingLevyskewalpha} shows how
a continuous distribution is confined and discretised to a probability
mass function on $1\bigstar,\dots,10\bigstar$.

The question is now: What is this distribution and how universal can it
be parameterised? Before trying to answer this question by looking at the
data we formulate a simple social theory which limits the
possible distributions to 'Gaussian-like' shapes.

It is natural to assume that people make their mind about a movie not
independent of the opinions of others. Each cinemagoer might adjust her
initial impression towards the opinions of others, towards the existing
mean rating or towards ratings of professional critics. This is modelled
by taking an average of several opinions as the final opinion of a
cinemagoer. Here, several aspects might be important like social networks
including correlations of links and initial impressions, opinion leaders,
timing effects and so on. But if we assume that initial impressions are
drawn from a random variable with finite variance, averaging of a large
enough number of opinions implies a distribution of averaged opinions
close to a normal distribution due to the central limit theorem. This
holds also when individual random variables are different under some
additional mild assumptions. Also for contrasting forces like 'if I
observe the average to be $1\bigstar$ higher then my opinion, I lower my
opinion $1\bigstar$' the limit theorem holds, as long as the forces are
linear. According to this theory of opinion making the histogram of
ratings should be a discretised and confined probability density function
of a normal distribution. The normal
distribution does not fit well, as it will turn out. Either the highest peak is not
achieved or the decay of bin size with distance from the highest peak is
too fast.

Alternatively, we might assume, that initial impressions are drawn from
fat-tailed distributions. This implies that distributions do not
have a finite variance. The probability of extreme initial impressions might not
vanish exponentially but as a power law with exponent $-(1+\alpha)$. If
this is the case a generalisation of the central limit theorem says that
an average of these random variables has a distribution close to a L\'evy
skew $\alpha$-stable distribution (the parameter $\alpha$ must indeed be
universal for this theorem). So, we can keep the theory of averaging, but
extend from the normal distribution to the wider class of L\'evy skew
$\alpha$-stable distributions.

\begin{figure*}
\includegraphics[width=0.32\textwidth]{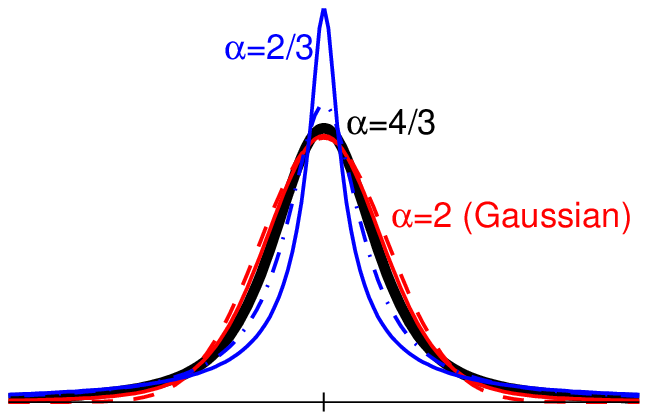}
\includegraphics[width=0.32\textwidth]{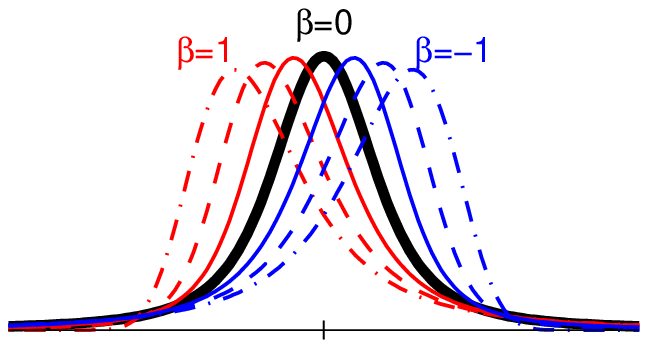}
\includegraphics[width=0.32\textwidth]{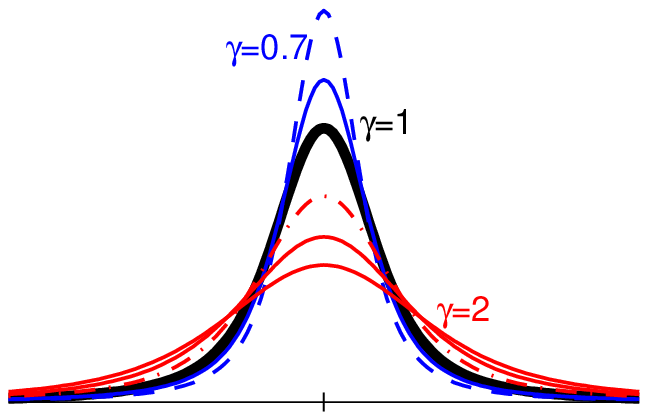}
\caption{Role of parameters $\alpha, \beta, \gamma$ for the shape of the probability density function of L\'evy skew $\alpha$-stable distributions. The base line case in black is the probability density function of $S(\frac{4}{3},0,1,\mu)$ with mu marked on the x-axis. } \label{fig:explevy}
\end{figure*}

The L\'evy skew alpha-stable distributions are the only stable
distributions (see \cite{Nolan2010StableDistributions}). It has four
parameters $\alpha, \beta, \gamma, \mu$ and is abbreviated
$S(\alpha,\beta,\gamma,\mu)$. (There are several parametrisations
  of the L\'evy skew $\alpha$-stable distribution. The one used here is
  $S(\alpha,\beta,\gamma,\mu;1)$ as explained in
  \cite{Nolan2010StableDistributions}.) Its probability density function
is
\begin{equation}\label{eq:1}
  f_{S(\alpha,\beta,\gamma,\mu)}(x) = \frac{1}{2 \pi} \int_{-\infty}^{+\infty}
\varphi (t;\alpha,\beta,\gamma,\mu)
  e^{-itx}\,dt
\end{equation}
with $\varphi(t;\alpha,\beta,\gamma,\mu)$ being its characteristic
function given by
\begin{equation}\label{eq:2}
  \varphi(t;\alpha,\beta,\gamma,\mu) = 
  \exp\left[\mu\!-\!|\gamma t|^\alpha\,(1\!-\!i
\beta\,\textrm{sign}(t)\Phi)~\right].
\end{equation}
and $\Phi = \tan(\frac{\pi\alpha}{2})$ if $\alpha\neq 1$ and $\Phi =
-\frac{2}{\pi}\log |t|$. The four parameters are
$\alpha\in]0,2]$, , $\beta [-1,1]$, $\gamma \in [0,\infty[$, and
$\mu\in]-\infty,\infty[$. The first two parameters are shape parameters,
where $\alpha$ represents the peakedness and $\beta$ the
skewness; $\mu$ and $\gamma$ are location and scale parameters. 
(But notice that $\beta$ is not the skewness in terms of the third moment, 
and $\alpha$ is not the peakedness in terms of kurtosis.) For $\alpha>1$,
$\mu$ also represents the mean of the distribution (otherwise not
defined). Figure \ref{fig:explevy} shows how the parameters modify the shape of the probability density function. 
Small $\alpha$ represents a
sharp peak but heavy tails which asymptotically decay as power laws with
exponent $-(\alpha+1)$. Maximal $\alpha=2$ is the normal distribution
with exponential decay at the tails. Scale parameter $\gamma$ corresponds to
the variance
$\sigma^2$ by the relation $\sigma^2 = 2\gamma^2$ only for $\alpha
=2$. For lower $\alpha$ the variance is infinite. Skewness $\beta=0$ gives a
distribution symmetric around the mean, a positive $\beta$ implies a
heavier left tail, a negative $\beta$ a heavier right tail, but with the
same decay on both sides. Only in the case $\beta=\pm 1$ one tail vanishes
completely. If $\alpha=2$ then $\beta$ has no effect. Only the special
cases of the normal distribution ($S(2,0,\sigma,\mu)$), the Cauchy
distribution ($S(1,0,\gamma,\delta)$) and the L\'evy distribution
($S(\frac{1}{2},1,\gamma,\delta)$) have closed form expressions.

In the following the L\'evy skew $\alpha$-stable distributions
(discretised and confined) will be fitted for each empirical rating
distributions in the data set.

\section{Fitted L\'evy skew $\alpha$-stable distributions}

Fitting is done by minimising least squares of the difference of the
normalised empirical rating histogram to the confined and discretised
probability density function of L\'evy skew $\alpha$-stable distributions
with parameters $(\alpha,\beta,\gamma,\mu)$. (For numerical
  reasons, fitting has been done with a different parameterisation
  $S(\alpha,\beta,\gamma,\delta;0)$ (see
  \cite{Nolan2010StableDistributions}). The parameters
  $\alpha,\beta,\gamma$ are equal to the former parameterisation and $\mu
  = \delta - \beta\gamma\tan(\pi\alpha/2)$.) Computation was performed
as follows: The values of the probability density function
$f_{S(\alpha,\beta,\gamma,\mu)}(x)$ are computed for $x=-20,-19,\dots,29,30$
by computing and integrating the characteristic function (Eq. \ref{eq:2}) on
$t=-20.005,\stackrel{+0.01}{\dots}:20.005$. Then values for $x=-20,\dots,1$ are summed
up and set on bin 1 and values for $x=10,\dots,30$ are summed up and
set on bin 10. This produces a probability mass function on $1,\dots,10$ for
$(\alpha,\beta,\gamma,\mu)$. Results were reasonably good, the missing
mass of the tails (below -20 and above +30) was mostly below $0.3\%$. The
fitting was computed by minimising the squares of distances of the
probability mass function for $S(\alpha,\beta,\gamma,\mu)$ to the
normalised empirical rating distribution. The minima were found with the
\texttt{matlab}-function \texttt{fminsearch}. The search converged in
$1081$ cases ($99.5\%$) the remaining cases it terminated by maximum
number of iterations. Finding a global minimum is not guaranteed by this
method, but results looked convincing. (Experimentally, some fits
  have been computed via minimising by gradient descent. It lead to very
  similar fits.) We refer to this fit as
$\textrm{fit}(\alpha,\beta,\gamma,\mu)$ Examples of
fits are shown in Figure \ref{fig:figExplainRatingLevyskewalpha}.

Table \ref{tab:fit} shows the mean values of fitted parameters over all
movies as well as goodness-of-fit measures. The sum of squared
error ($\textrm{SSE} = \sum_{i=1}^{10}(r_i -
  f_{S(\alpha,\beta,\gamma,\mu)}(i))^2$ with $r_i$ being the fraction of
  ratings for $i\bigstar$) is on average very small, the
coefficient of determination $R^2$
is on average almost one. ($R^2 =
  1-\frac{\textrm{SSE}}{\sum_{i=1}^{10}(r_i-\mean{r})}$ with $r_i$ the
  fraction of ratings for $i\bigstar$ (therefore $\mean{r}=0.1$).)
Both reflects that indeed most fits also look
impressively close to the empirical histogram. Further on, a
Kolmogorov-Smirnov test has been performed for each movie. (Done with the
  \texttt{matlab}-function \texttt{kstest2} on the vector of all ratings
  and a vector with the same number of ratings as expected according to
  the fit.) With level of significance $0.05$ the null hypothesis that
the expected fitted distribution and the empirical histogram are drawn
from the same distribution could not be rejected for $68.7\%$ of the
movies. The Kolmogorov-Smirnov test is very hard, it rejects the null
hypothesis very likely for large samplesizes. Given the high number of ratings ($>20,000$) for
each movie this rate is still impressive. But it is also clear that L\'evy skew $\alpha$-stable
 cannot fully explain all possible rating histograms.

For comparison Table \ref{tab:fit} also contains mean values for a fit
with normal distributions $S(2,0,\gamma,\mu)$. The goodness-of-fit
parameters are worse. This is natural because there are less free
parameters, but clearly the normal distribution is ruled out as an
appropriate candidate.

\begin{table*}[htbp]
  \centering
  \begin{tabular}[t]{r|r||r||r|r|r|l}
    & $\mean{\textrm{ratings}}$        & corr-coef &
$\mean{\textrm{fit}(\alpha,\beta,\gamma,\mu)}$ &
$\mean{\textrm{fit}(2,0,\gamma,\mu)}$ & $\mean{\textrm{fit}(\mu)}$ & \\ \hline
    mean         & 7.3464  & 0.5772  & 7.6326 & 7.6590 & 7.5862       & $\mu$   
                 \\ 
    std          & 1.9669  & 0.8883  & 1.2021 & 1.2456 &  1.1993      & $\gamma$
                 \\
    skewness     & -1.0610 & -0.2829 & 0.0159 & 0      & -0.0114      & $\beta$ 
                 \\
    kurtosis     & 1.8581  & -0.0138 & 1.3261 & 2      & $\frac{4}{3}$& $\alpha$
                 \\ \hline
    \multicolumn{3}{c|}{}            & 0.0002 & 0.0035 & 0.0035       & 
$\textrm{SSE}$    \\
    \multicolumn{3}{c|}{}            & 0.9965 & 0.9404 & 0.9434       & 
 $R^2$             \\
    \multicolumn{3}{c|}{}            & 68.7\% & 0\%    & 4.5\%        &  K-S
$\checkmark$         \\ \cline{4-7}
  \end{tabular}
  \caption{Aggregated measures on the data set and on three
confined L\'evy skew $\alpha$-stable fits
($\textrm{fit}(\alpha,\beta,\gamma,\mu)$, $\textrm{fit}(2,0,\gamma,\mu)$,
$\textrm{fit}(\mu)$). Mean, standard deviation, skewness, and kurtosis are
computed for the histograms of each movie. Parameters of fits are also computed
for each movie. The table shows the mean values for all 1,086 movies. The
correlation coefficient is computed for the `analog' measures for the ratings
and $\textrm{fit}(\alpha,\beta,\gamma,\mu)$. The low (and negative) correlation
skewness vs. $\beta$ and kurtosis vs. $\alpha$ show that the parameters of the
fit deliver information on the distribution which is not extracted by the
`standard measures' on the raw data. Goodness-of-fit parameter are computed for
each fit for each movie. The mean over all movies is shown for sum of squared
error (SSE) and coefficient of determination R-square. For the
Kolomogorov-Smirnov test (K-S) the rate of not rejecting of the hypothesis of a
common distribution of ratings and fitted distribution is given.}
  \label{tab:fit}
\end{table*}

\begin{figure*}[htbp]
  \centering
  \includegraphics[width=0.4\textwidth]{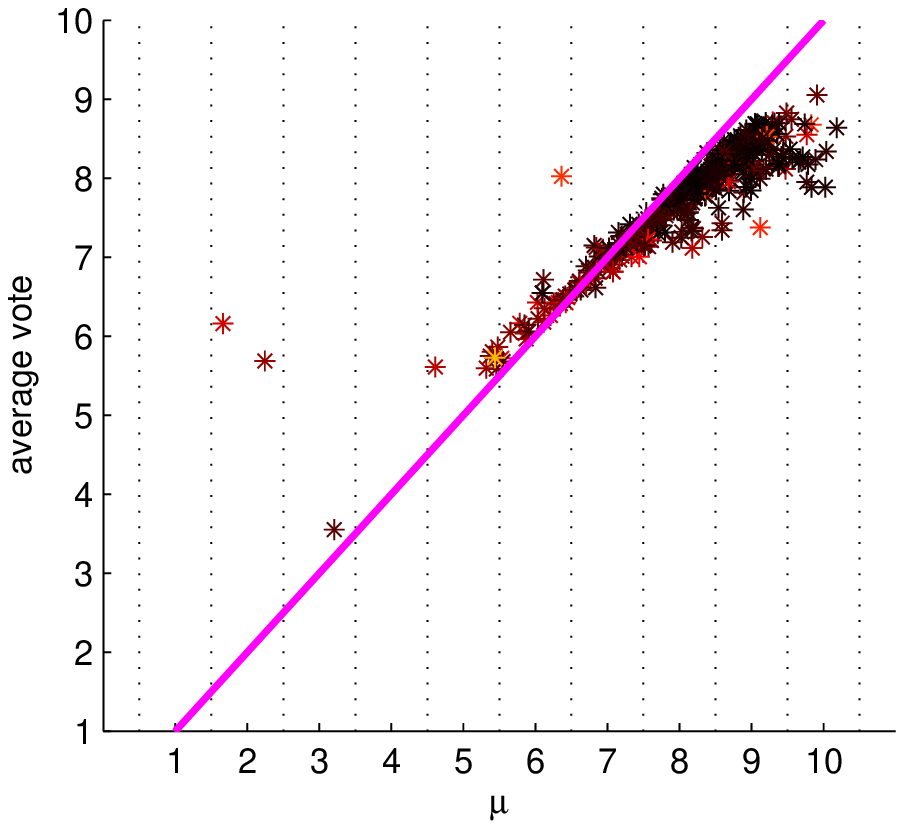}
  \includegraphics[width=0.4\textwidth]{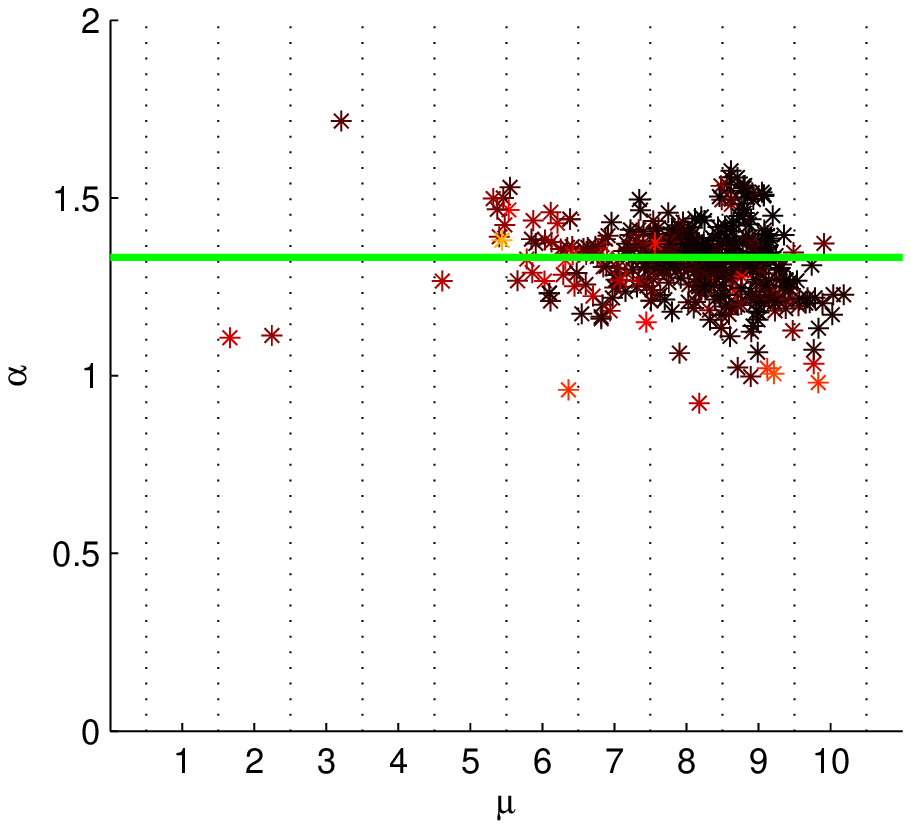}
  \includegraphics[width=0.4\textwidth]{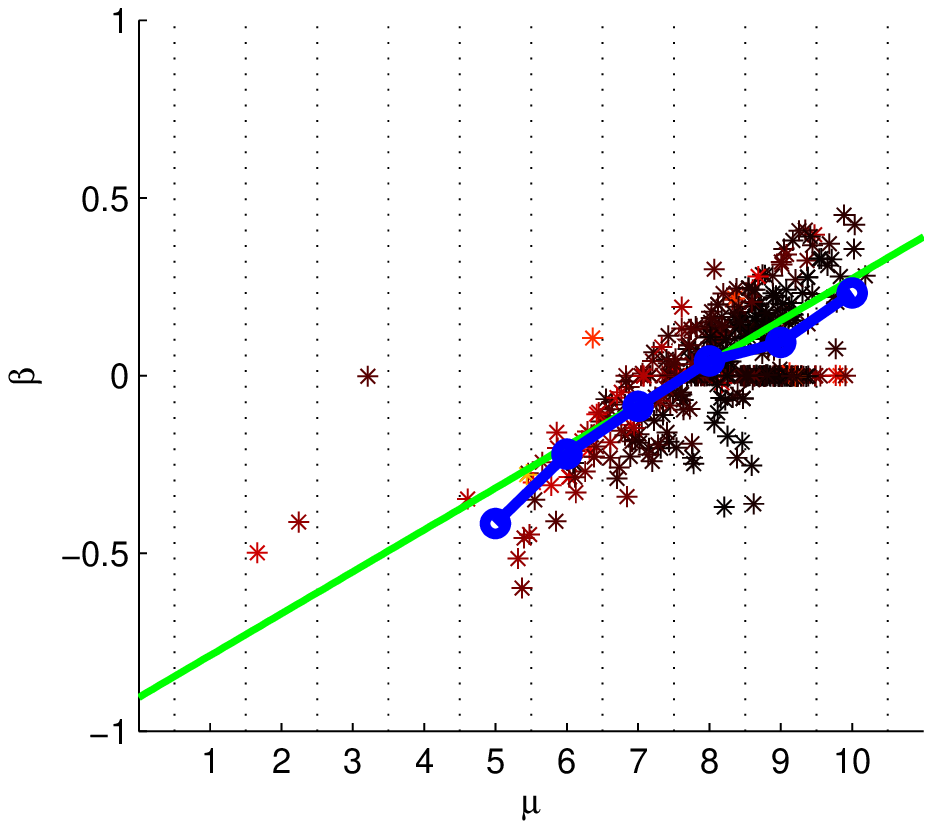}
  \includegraphics[width=0.4\textwidth]{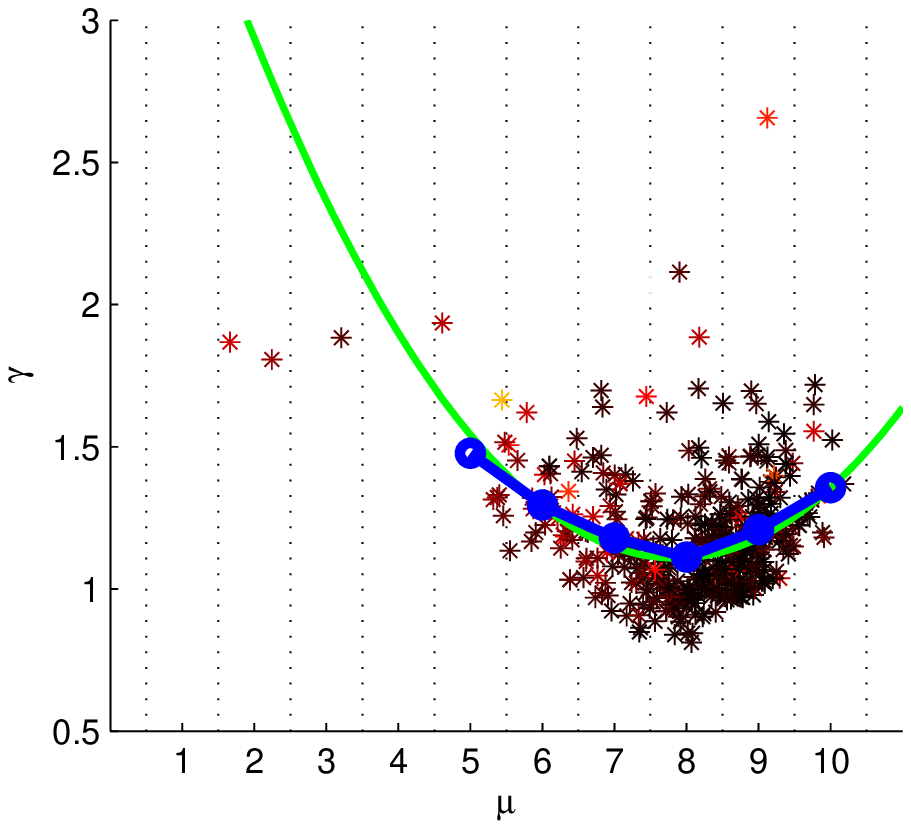}
  \caption{Parameters of best fit for confined Levy skew
    $\alpha$-stable distributions for all movies. $\mu$ is the mean
    of the distribution, $\alpha$ modifies peakedness and tail exponent,
    $\beta$ skewness, and $\gamma$ scales how broad the distribution is.}
  \label{fig:figLevyParams}
\end{figure*}

Figure \ref{fig:figLevyParams} shows the parameters of best fits
for all movies as scatter plots. All four subplots show $\mu$ at the
abscissa. Dark points indicate movies which fits have a small sum of
squared errors (SSE), red stars indicate medium SSE, and yellow stars
indicate bad fits with high SSE.

The first plot shows $\mu$ with respect to the original average of
ratings. It shows that $\mu$ is spread wider than the original
average. So, $\mu$ can serve as a measure for movie quality which
differentiates better than the original average.

The remaining three subplots show the relations of $\mu$ to the other
three parameters $\alpha,\beta,\gamma$ of the best L\'evy skew
$\alpha$-stable fits. The blue dots in the two bottom plots show the
averages of the ordinate values within the $\mu$-region marked by the
grid lines. The green lines represents $\alpha=\frac{4}{3}$, for $\beta$
the best linear fit for the blue dots, and the best quadratic fit for
$\gamma$. The plots show that the peakedness $\alpha$ concentrates to
values between $1.2$ and $1.5$, which is clearly not normal
distributed. The average value is $\mean{\alpha} = 1.3261\approx
\frac{4}{3}$.  For the skewness $\beta$ there is a clear trend with
respect to $\mu$. Interestingly, $\beta=0$ is most likely almost exactly
at $\mu = 7.6326$ which is equal to $\mean{\mu}$. For better movies there is an additional
positive skewness (meaning that the right tail is fatter). Respectively,
for movies worse than $\mean{\mu}$ there is additional negative skewness
(meaning that the left tail is fatter). For the scale parameter there is
also a clear trend visible. The most narrow distribution is achieved also
almost exactly for movies with $\mu = \mean{\mu}$. For better
and worse movies the distributions get broader.

It is not apriori clear and thus remarkable that $\mean{\mu}$ plays a central role for the
deviations in $\beta$ and $\gamma$ with respect to $\mu$. This gives rise
to the speculation that $\mean{\mu}$ is kind of universal modulo the
scale of ratings (here $1,\dots,10$). This is underpinned by the finding
of \citet{Cosley.Lam.ea2003Isseeingbelieving} that users rate
consistently in different rating schemes. If we rescale $\mean{\mu}=7.6326$ to the scale $1,\dots,5$ we get
$4.0633$ which coincides almost exactly with $4.07$ which is the average
mean rating of books averaged over all books in the
\texttt{amazon.com}-sample of \citet{Li.Hitt2004SelfSelectionand}. Rescale
  is done under the assumption that each rating stands for a bin centred
  on the rating with width equal to the distance of successive ratings
  (here 1). Thus a $10\bigstar$-rating $r_{10}$ is converted to
  the 5$\bigstar$-rating by $r_{5} =
  5(\frac{r_{10}-0.5}{10})+0.5$. This ensures for examples that
  $1\bigstar$ in a $5\bigstar$-rating corresponds to $1.5\bigstar$ stars
  in a 10$\bigstar$-rating, respectively $5\bigstar$ corresponds to
  $9.5\bigstar$. It
does not coincide as good with $3.44$ which was found by
\cite{Dellarocas.Narayan2006StatisticalMeasureof} for
\texttt{movies.yahoo.com}-data. The deviation may come from two differences: First, in \cite{Li.Hitt2004SelfSelectionand} and in this study the average reported is the average of the average ratings of movies, while \cite{Dellarocas.Narayan2006StatisticalMeasureof} reports the the pure average rating over all ratings in the database. Second, \cite{Li.Hitt2004SelfSelectionand} and this study select books respectively movies similar: this study by all having more than 20,000 ratings, and \cite{Li.Hitt2004SelfSelectionand} by being on a bestseller list and having a sufficient number of reviews. Both sampling method imply a similar selection bias which is different from \cite{Dellarocas2003DigitizationofWord} which collects all movies released in 2002.

Taking this speculation as true it means that an average movie receives
an average vote of about $0.71$ on a generalised scale $[0,1]$. This
indicates a universal strong positive bias for the average
movie. The strong positive bias may be implied by an overall
  selection-bias, that user select movies or products they are likely to
  like or even they like movies and products just because they paid for
  them. Contrasting, a negative bias is reported on ratings for jokes in 
  \cite{Goldberg.Roeder.ea2001EigentasteConstantTime}. Following the results of
$\textrm{fit}(\alpha,\beta,\gamma,\mu)$ we further conclude that the
distribution of ratings for an average movie has no skewness ($\beta=0$)
and the smallest scale parameter (here $\gamma=1.1$). If a movie deviates
from average this implies higher deviations in the distribution
($\gamma>1.1$) and a skewness which pronounces the deviation from the
average movie. The latter observation can be regarded as a hint for a
socially implied positive feedback in determining opinions on movies
which quality is above (or below) an average movie.

Taking the trends displayed by the green lines in Figure
\ref{fig:figLevyParams} one can construct a one-parameter fit on $\mu$
with $\alpha=\frac{4}{3}$, $\beta$ determined by the linear fit and
$\gamma$ by the quadratic fit. The equations to compute
  $\beta,\gamma$ from $\mu$ are $\beta = b_1\mu + b_2$ and $\gamma =
  c_1\mu^2+c_2\mu+c_3$ with parameters $b_1=0.1178, b_2=-0.9049, c_1 =
  0.05342, c_2 = -0.8388, c_3 = 4.401$. We refer to this fit as
$\textrm{fit}(\mu)$. Mean values and mean goodness-of-fit measures are
also shown in Table \ref{tab:fit}. The one-parameter fit gets better
goodness than the two parameter $\textrm{fit}(2,0,\gamma,\mu)$.

\begin{figure*}[htbp]
  \centering
  \includegraphics[width=0.3\textwidth]{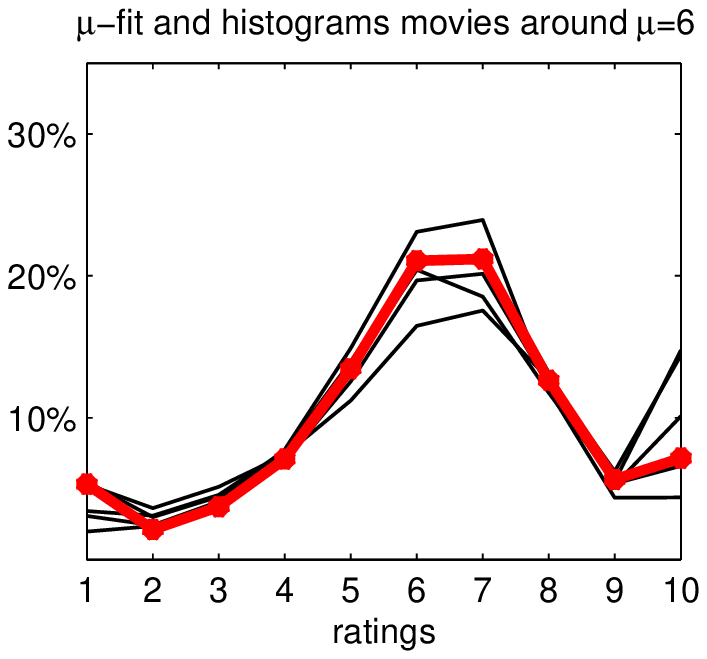}
  \includegraphics[width=0.3\textwidth]{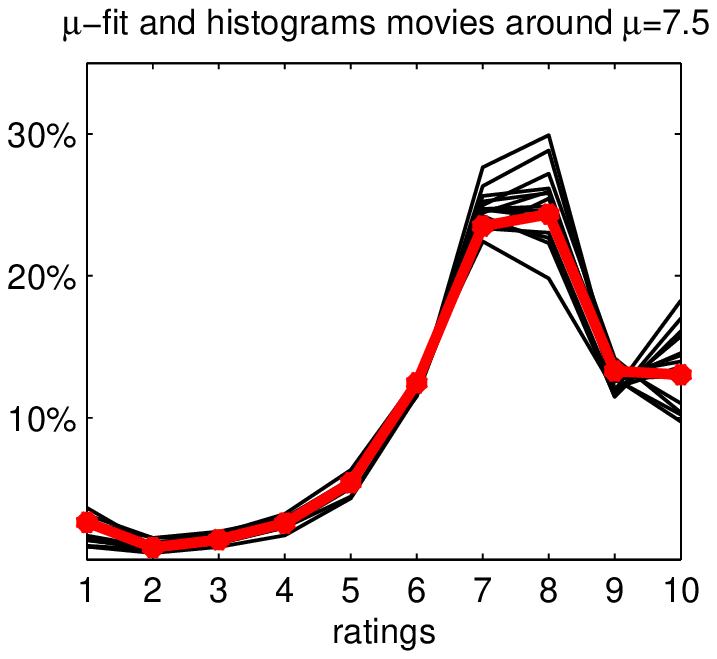}
  \includegraphics[width=0.3\textwidth]{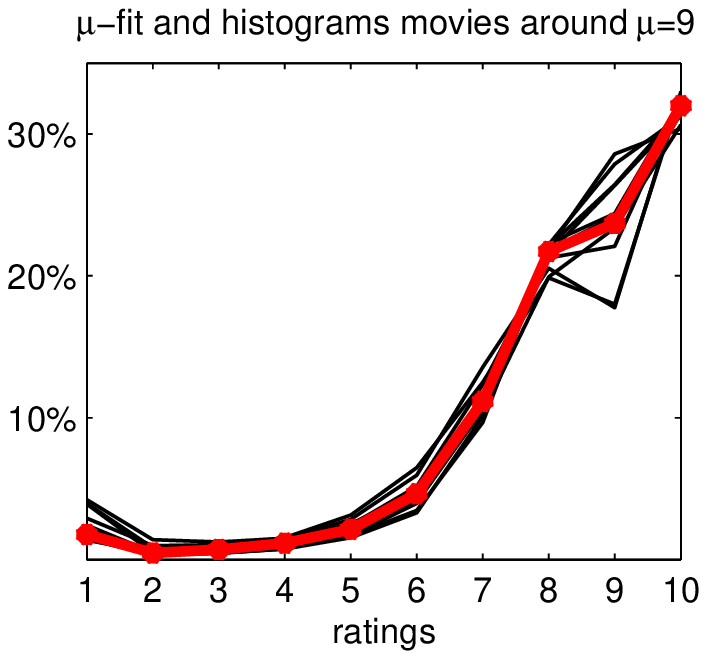}
  \caption{The theoretical probability mass functions
    according to the parameters of $\textrm{fit}(\mu)$ for $\mu=6,7.5,9$
    and all empirical histograms which received best fitted values for
    $\mu \in [5.98,6.02], [7.48,7.52], [8.98,9.02]$. }
  \label{fig:figOneParamExample}
\end{figure*}

Figure~\ref{fig:figOneParamExample} shows how $\textrm{fit}(\mu)$ is able
to approximate empirical histograms. The shape of empirical distributions
is well captured but variations for different movies are big enough to
conclude that $\textrm{fit}(\mu)$ can only be seen as a baseline
case. Movies can have some individual characteristics of their rating
distribution which go beyond the quality (captured in
$\mu$). Deviations from the baseline case can be used to classify movies
in a new way to understand what the cause of deviations might
be. This is a task for further research.

\begin{figure*}[htbp]
  \centering
  \includegraphics[width=0.35\textwidth]{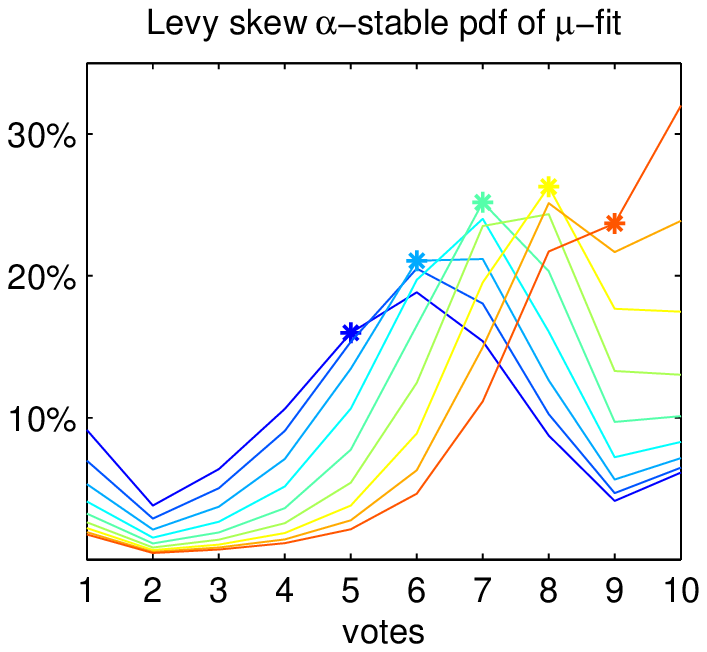}
  \includegraphics[width=0.35\textwidth]{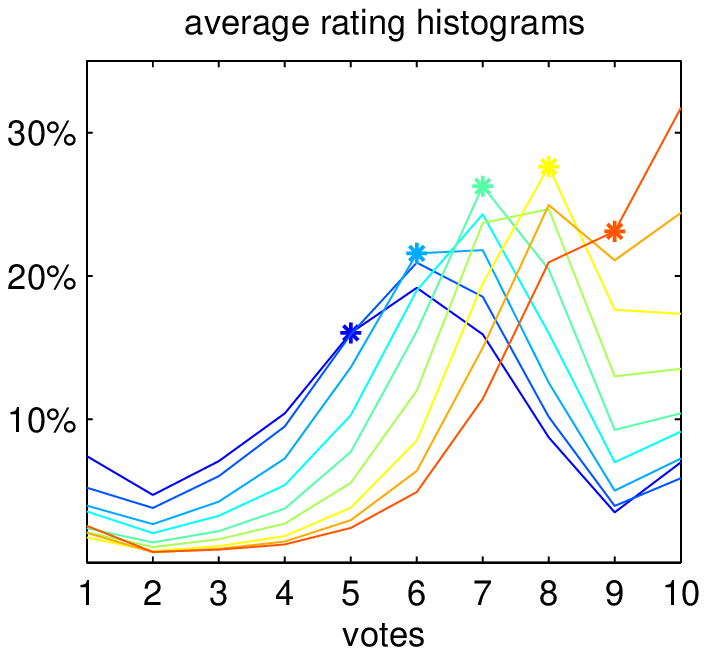}
  \includegraphics[width=0.35\textwidth]{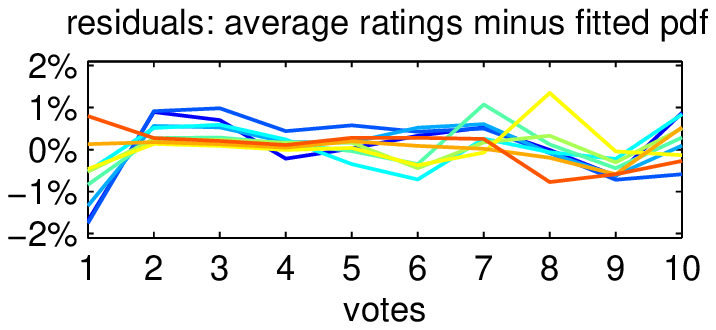}
  \caption{The theoretical probability mass functions
    according to the parameters of $\textrm{fit}(\mu)$ for $\mu =
    5,\stackrel{+0.5}{\dots},9$ (left), empirical average histograms for
    all movies with best fitted values for $\mu \in [4.75,5.25]$,
    $[5.25,5.75]$,$\dots$, $[8.25,8.75]$, $[8.75,9.25]$ (right), and
    residuals of both.  The stars mark the underlying $\mu$-values of
    curves. Colors are the same in all plots.}
  \label{fig:figOneParameterFit}
\end{figure*}

Finally, Figure \ref{fig:figOneParameterFit} shows a comparison of
theoretical histograms of $\textrm{fit}(\mu)$ and the average empirical
histograms. The theoretical histograms are for $\mu =
5,\stackrel{+0.5}{\dots},9$ and the average empirical histograms are over
all movies with fitted value of $\mu$ within the intervals $\mu \in
[4.75,5.25]$, $[5.25,5.75]$, $\dots$, $[8.25,8.75]$, $[8.75,9.25]$. The
similarity underpins that $\textrm{fit}(\mu)$ can really serve as a good
baseline case. But some deviations from the baseline case seem to be not
totally random. E.g. the residuals show that the size of the $1\bigstar$
bin for low quality movies ($\mu<7$) is on average predicted too high,
while the $2\bigstar,3\bigstar$ bins are on average predicted too low.

\section{Conclusion}

With some success rating histograms were fitted to confined L\'evy skew
$\alpha$-stable
distributions. This clearly demonstrates that the assumptions that
opinions are normally distributed, beta distributed or U-shaped around the quality of the movie is not
valid. Some histograms have of course a U-shaped (or better J-shaped) form, e.g. right-hand side in Figure \ref{fig:figExplainRatingLevyskewalpha}. But a U-shape can not approxiamte all histograms, e.g. left-hand side of Figure \ref{fig:figExplainRatingLevyskewalpha}. 

If the assumption that expressed opinions of users are weighted
averages of formerly expressed opinions of others this implies that these
opinions must come from distributions with fat tails with a power-law
exponent of about $1.2$ to $1.5$ to get good fits. Further on, the scale
and skewness parameter of the best fits change systematically with the
deviation of its mean from the mean of an average movie (with $\mu
\approx 7.6$). A movie better than average shows right skewness and a
larger scale parameter. A movie worse than average shows left skewness
and a also a larger scale parameter. Thus, better movies have also a
heavier tail on the better side and worse movies have a heavier tail on
the worse side. In general, distributions get broader when deviating from
the mean. Both observations seem plausible from a sociological point of
view.  The new measures of skewness ($\beta$) and peakedness ($\alpha$)
are not the same as the classical skewness and excess kurtosis which are 
compute directly from the sample data (see Table \ref{tab:fit}). There is
no correlation of both measures, or even a negative one. This underpins,
that fitting rating histograms as confined distributions really delivers
a new characterisation. A further advantage of this approach is, that the
L\'evy skew $\alpha$-stable distribution defines a distribution completely,
which mean, standard deviation, skewness and excess kurtosis do not. 

A one-parameter fit based on this observations shows to approximate the
data well, but is not able to establish a strict characterisation of
movie histograms. Deviations from the constructed baseline case are not
neglectable. Nevertheless, it could be useful to characterise movies by
their deviation from their baseline case. Further on, there might is a selection bias in the data, because only movies with a large number of ratings were selected. The fit might not work for less rated movies. The method might be used to
detect attacks of enthusiastic fans (or movie companies) which try to
rate movies up.

There seems to be some universality in movie rating distributions, which
may be implied by people adjusting their opinions with peers and other
sources of opinions. Clearly, other theories which may imply other
underlying distributions need to be developed and checked against data and
also this theory needs to be checked against data from other sources to
clarify universality in continuous opinion dynamics about taste.

\paragraph{Acknowledgement} The research leading to these results has received funding from the European Community's Seventh Framework Programme (FP7/2007-2013) under grant agreement no. 231323 (CyberEmotions project).

\bibliographystyle{epj}

\end{document}